\documentclass[lettersize,journal]{IEEEtran}
\usepackage{amsmath,amssymb,amsfonts}
\usepackage{algorithmic}
\usepackage{algorithm}
\usepackage{array}
\usepackage[caption=false,font=normalsize,labelfont=sf,textfont=sf]{subfig}
\usepackage{textcomp}
\usepackage{stfloats}
\usepackage{url}
\usepackage{verbatim}
\usepackage{graphicx}
\usepackage{cite}
\usepackage[table]{xcolor}
\usepackage{array}
\usepackage{soul,color}
\usepackage{subcaption}
\usepackage{changepage}
\usepackage[font=small,labelfont=bf]{caption}
\usepackage{caption}
\usepackage{afterpage}
\usepackage{multirow}
\usepackage{adjustbox}
\usepackage{booktabs}
\usepackage{lipsum}
\usepackage{titlesec}
\usepackage{tabularx}
\usepackage{xcolor}
\usepackage{soul} % For strikethrough text
\usepackage[normalem]{ulem} % For strikethrough

\begin{document}

\title{A Multi-layer Non-Terrestrial Networks Architecture for 6G and Beyond under Realistic Conditions and with Practical Limitations}

\author{Faical Khennoufa,
Khelil Abdellatif,
Halim Yanikomeroglu,  
Metin Ozturk, \\ Taissir Elganimi,
Ferdi Kara, and Khaled Rabie
        % <-this % stops a space
\thanks{ F. Khennoufa and  A. Khelil are with LGEERE Laboratory, Department of Electrical Engineering, Echahid Hamma Lakhdar University, El-Oued, Algeria, H. Yanikomeroglu is with Non-Terrestrial Network (NTN) Laboratory, Department of Systems and Computer Engineering, Carleton University, Ottawa, K1S 5B6, ON, Canada. M. Ozturk is with Electrical and Electronics Engineering, Ankara Yıldırım Beyazıt University, Ankara, Turkiye, and Non-Terrestrial Network (NTN) Laboratory, Department of Systems and Computer Engineering, Carleton University, Ottawa, K1S 5B6, ON, Canada. T. Elganimi is with the Department of Electrical and Electronic Engineering, University of Tripoli, Libya. F. Kara is with the Department of Computer Engineering, Zonguldak Bulent Ecevit University, Zonguldak, Turkey, and Division of Communications Systems, KTH Royal Institute of Technology, Sweden. K. Rabie is with the Department of Computer Engineering, King Fahd University of Petroleum and Minerals, Saudi Arabia.

}% <-this % stops a space
}

\markboth{IEEE Internet of Things Magazine }%
{Shell \MakeLowercase{\textit{et al.}}: A Sample Article Using IEEEtran.cls for IEEE Journals}

\maketitle

\begin{abstract}
In order to bolster the next generation of wireless networks, there has been a great deal of interest in non-terrestrial networks (NTN), including satellites, high altitude platform stations (HAPS), and uncrewed aerial vehicles (UAV). To unlock their full potential, these platforms can integrate advanced technologies such as reconfigurable intelligent surfaces~(RIS) and next-generation multiple access (NGMA). 
However, in practical applications, transceivers often suffer from radio frequency (RF) impairments, which limit system performance.
In this regard, this paper explores the potential of multi-layer NTN architecture to mitigate path propagation loss and improve network performance under hardware impairment limitations.
First, we present current research activities in the NTN framework, including RIS, multiple access technologies, and hardware impairments.
Next, we introduce a multi-layer NTN architecture with hardware limitations.
This architecture includes HAPS super-macro base stations (HAPS-SMBS), UAVs--equipped with passive or active transmissive RIS--, and NGMA techniques, like non-orthogonal multiple access (NOMA), as the multiple access techniques to serve terrestrial devices.
Additionally, we present and discuss potential use cases of the proposed multi-layer architecture considering hardware impairments. The multi-layer NTN architecture combined with advanced technologies, such as RIS and NGMA, demonstrates promising results; however, the performance degradation is attributed to RF impairments. Finally, we identify future research directions, including RF impairment mitigation, UAV power management, and antenna designs.
\end{abstract}

\section*{Introduction}
The sixth-generation (6G) technology standard entails significant changes in network structures and technologies. 
Particularly, non-terrestrial networks (NTN) integrated with 6G networks offer a cost-effective approach to establishing seamless and expansive wireless connectivity across diverse environments—including rural, remote, and urban areas \cite{alfattani2021link}.
According to the Third Generation Partnership Project (3GPP), NTN includes three main kinds of flying objects, namely satellites, high altitude platform stations (HAPS), and uncrewed aerial vehicles (UAV) \cite{3gpp2018study}. 
NTN can act as an extension to the current terrestrial network, increasing its capacity and extending coverage, which is consistent with the usage scenarios and capabilities described by the International Telecommunication Union (ITU) vision document for International Mobile Telecommunications (IMT)-2030~\cite{ITU2023}.

A HAPS is a network node that operates in the stratosphere, providing flexibility, wide connectivity, security (due to its high altitude positioning, which makes it less vulnerable to ground threats), and reliability.
The integration of UAVs with HAPS has been proposed due to its many advantages, such as high mobility, on-demand deployment, low path-loss, and high probability of establishing a line-of-sight (LoS) link with terrestrial devices. 
UAVs can improve the system performance in situations, such as poor network coverage or crowded areas. 
In order to improve the system performance, several technologies are integrated with NTN, including reconfigurable intelligent surfaces (RIS), multiple antennas, and next-generation multiple access (NGMA) like non-orthogonal multiple access (NOMA), rate-splitting multiple access (RSMA), and spatial division multiple access (SDMA) \cite{abbasi2024hemispherical,alfattani2021link,kurt2021vision}.

On the other hand, over the last decade, a great amount of effort has been devoted to developing integrated, cost-effective, and energy-efficient radio frequency (RF) transceivers. However, these systems suffer from inevitable RF front-end impairments due to component mismatches and manufacturing defects. 
Despite the impact of the RF front-end on system performance, its detrimental effect is often overlooked and has not yet been explored in the context of NTN. 
Therefore, there is a call for a comprehensive investigation into this topic, which is crucial to realistically determine the performance limits of NTN communications in terms of spectral efficiency, energy efficiency, and other performance metrics. 
Motivated by this, this paper aims to highlight the potential of multi-layer NTNs for 6G and beyond in the presence of hardware impairments. Advanced technologies like multiple antennas, RIS, NGMA, etc., have been discussed separately with NTN in the literature. Therefore, this paper discusses the integration of these technologies with multi-layer NTNs to highlight their performance benefits in complex environments. The comparative case study is conducted between multi-layer NTNs and several radio access technologies (RAT), which are considered under both ideal and non-ideal conditions, elaborating on performances achieved versus the constraints imposed by hardware.

\begin{figure*}[ht]
    \centering
    \includegraphics[width=\textwidth]{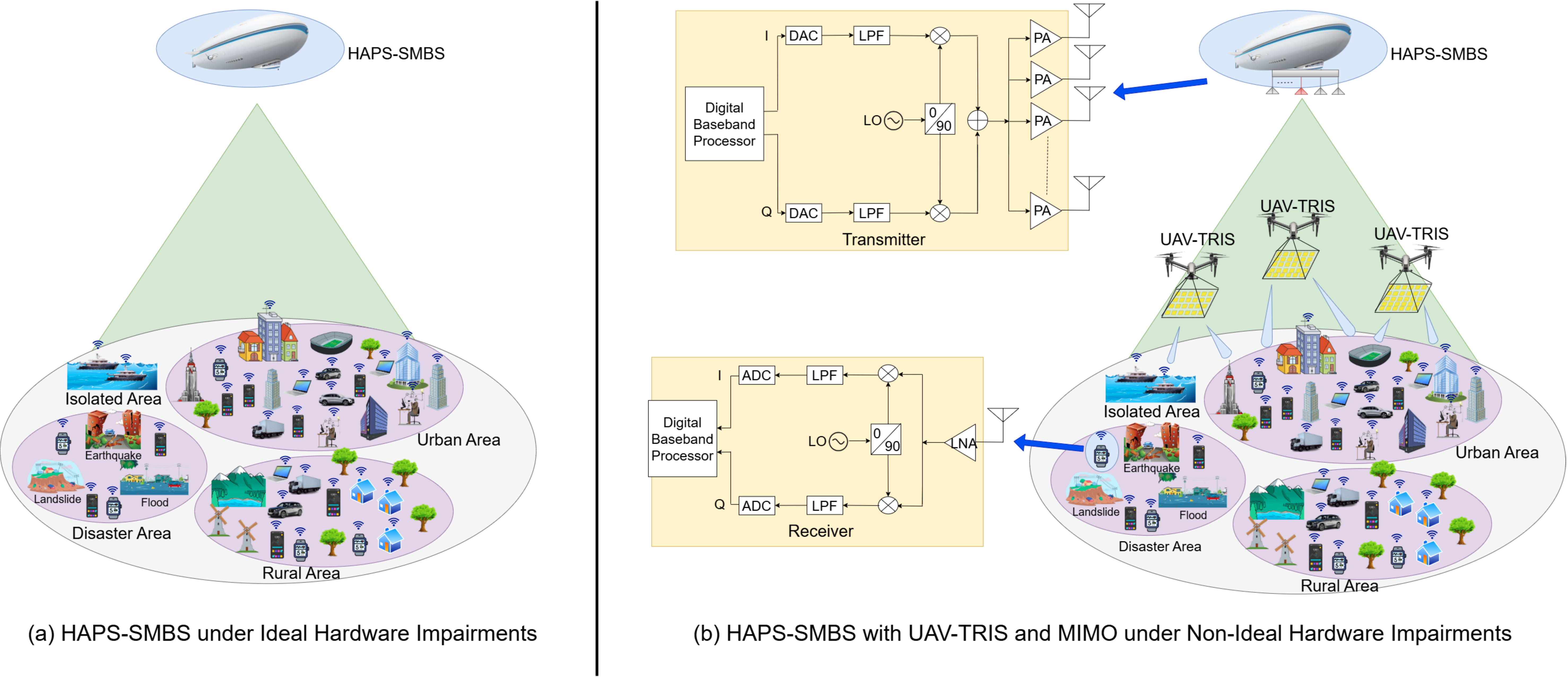}
    \caption{HAPS-SMBS assisted UAV-TRIS for terrestrial networks with hardware impairments.}
    \label{constellations}
\end{figure*}

\section*{Current Research Activities in NTN Framework}
The ITU document~\cite{ITU2023} aims to integrate terrestrial and NTN for 6G networks to improve connectivity, reduce latency, and provide high data rates in areas with insufficient ground infrastructure. NTN is classified into spaceborne and airborne vehicles. Spaceborne vehicles are satellites.
Airborne vehicles generally refer to drones or aircraft as part of a network's infrastructure, including uncrewed aircraft systems (UAS) platforms, HAPS, and UAVs. A new concept called HAPS as a super macro base station (HAPS-SMBS) or HAPS as an IMT base station (HIBS) is proposed in ITU documents~\cite{ITU2023,9356529}, which will play a key role in future networks.

In the context of NTN systems, integrating advanced technologies, such as RIS, MIMO, and NGMA (e.g., NOMA, RSMA, SDMA, etc), has attracted much attention. This integration aims to increase resource utilization efficiency, reduce path loss, and improve the received signal quality. The integration between NTN and these systems is described in the following subsections.

\subsection*{Reconfigurable Intelligent Surfaces} 
A RIS is a passive structure that can manipulate and forward signals in the desired direction, offering lower power consumption, wider coverage, and a higher data rate. RIS can be reflective, transmissive, or hybrid as described in~\cite{9365009}.
Moreover, active RIS with power amplification capabilities is recommended to address double path-loss and signal amplification issues~\cite{9998527}. 
Previous research has focused on the use of RIS in ground environments, e.g., equipping building facades with RIS as in~\cite{9365009,9998527}. From another perspective, the use of RIS in NTNs has recently been envisioned to improve signal propagation, coverage, and energy efficiency. RIS can be integrated on NTNs in various ways, such as coating the exterior of balloons or aircraft or installing it as a separate horizontal surface on the bottom, top, or sides of a UAV~\cite{alfattani2021link}. This integration provides signal control and redirection to difficult areas.

\subsection*{Multiple Antenna Technologies}
Multiple-input multiple-output (MIMO) technologies improve network capacity, reliability, and flexibility by using multiple antennas at the transmitter, receiver, or both ends. MIMO technology has the potential to increase the capacity and reliability of NTN. Given the operating costs and power constraints of NTN, antenna selection is an interesting way to reduce the cost and complexity of using multiple antennas~\cite{kurt2021vision,liu2022evolution}. In addition, the combination of NTN with massive MIMO technology, ultra-massive MIMO (UM-MIMO), and beamforming antenna technologies is among the pioneering and promising models for channel capacity and spectral efficiency enhancements~\cite{kurt2021vision,abbasi2024hemispherical}.

\subsection*{Multiple Access Technologies}
One of the major challenges facing NTNs is the efficient use of the limited spectrum available, making the design of an appropriate multiple access technology crucial to improving network capacity. 
NGMA techniques like NOMA, RSMA, and SDMA are able to address this challenge through better effective usage of spectrum and support a high number of users in complex network environments.
NOMA allows multiple users to share the same frequency bands without requiring additional bandwidth, thereby enhancing the performance and capacity of NTNs in the upcoming 6G landscape~\cite{kurt2021vision}.
By taking advantage of the spatial dimensions that multi-antenna NTN (e.g., HAPS) systems supply, SDMA assigns user equipment to each beam and employs superposition coding (SC) and successive interference cancellation (SIC) within each group. In the NTN, MIMO with NOMA (MIMO-NOMA) overloads SDMA by assigning different beams to clusters, enabling a larger number of users to share each beam, and promoting massive connectivity. MIMO-NOMA differs from multi-user MIMO in that the former allows a cluster of users, not just one user, to share one beam~\cite{kurt2021vision}.
Furthermore, to efficiently manage NTN traffic, RSMA is a promising multiple access strategy that separates device signals into common and private parts, optimizing spectrum usage and facilitating simultaneous data exchange~\cite{kurt2021vision,liu2022evolution}.

\section*{Multi-layer NTN Architecture with RAT and Hardware Impairments}
Current research on multi-layer NTN for serving terrestrial devices focuses on the assumption of ideal or perfect RF transceivers, as in~\cite{alfattani2021link,abbasi2024hemispherical,10387578}. 
On the practical side, RF transceivers suffer from different impairments that affect the overall performance of NTN platform communications. 
This section explores the capabilities of ideal multi-layer NTN technologies for terrestrial devices. Then,we discuss its capabilities under the technical limitations of RF transceivers with respect to overall performance.

\begin{table*}[h]
\centering
\caption{ Comparison of KPIs for different scenarios of multi-layer NTN architecture configurations~\cite{alfattani2021link, abbasi2024hemispherical,kurt2021vision,9356529,9365009,10387578,9579401}.}
\label{tab:compare}
\begin{tabular}{|c|c|c|c|c|}
\hline
\textbf{KPIs} & \textbf{HAPS-SMBS} & \textbf{HAPS-SMBS with UAV} & \textbf{HAPS-SMBS with UAV-TRIS} & \textbf{HAPS-SMBS with UAV-TRIS and MIMO} \\ 
\hline \hline
Coverage              & High      & High               & Very High               & Very High   \\ \hline 
Latency               & Medium    & Low                & Low                     & Very Low    \\ \hline
Signal Quality        & Medium    & High               & Very High               & Very High   \\ \hline
Mobility & High      & Very High          & Very High               & Very High   \\ \hline
Path Loss             & High      & Medium             & Low                     & Very Low    \\ \hline
Resilience           & Low       & Medium             & High                    & Very High   \\ \hline
Deployment Complexity & Low       & Medium             & High                    & Very High   \\ \hline
Operational Cost      & Low       & Medium             & High                    & Very High   \\ \hline
Scalability          & Medium    & High               & Very High               & Very High   \\ \hline
Security             & Medium    & High               & High                    & Very High   \\ \hline
Energy Efficiency     & Medium    & Medium             & High                    & Very High   \\ \hline
\end{tabular}
\end{table*}

\textbf{HAPS-SMBS with UAVs}: 
The HAPS-SMBS, operating at quasi-fixed altitudes up to 20 km, offers wide coverage, reliable, and low-latency communications in inaccessible regions. 
Nevertheless, they suffer from a variety of problems, including long distance and weather conditions. For example, direct communications between HAPS-SMBS and terrestrial networks, as shown in the scenario of Fig. 1 (a), have poor service quality due to line-of-sight loss caused by long distances and severe weather conditions. 
In this respect, the above limitations are effectively mitigated by the multi-layer NTN architecture exploiting complementary strengths from several platforms toward further improvement in network performance. In contrast to single-layer systems consisting of only HAPS-SMBS or only UAVs, the multi-layer architecture ensures dynamic and flexible networking through various layer strengths. For instance, wide-area coverage could be offered by HAPS-SMBS, while local area, on-demand connectivity in high-demand or hard-to-reach areas could be provided by UAVs. Thus, the multi-layer system provides a way for the network to adapt to different user requirements and environmental conditions, ensuring consistent and reliable communication.
Therefore, the integration of HAPS-SMBS and UAV can overcome direct connection issues and improve network capacity and coverage. UAVs can be strategically positioned to reduce the distance and path-loss in high-demand areas to offload traffic and enhance network performance~\cite{10387578}.
Furthermore, this interaction between layers, such as HAPS-SMBS to UAVs and UAVs to ground users, allows for efficient resource allocation and thus ensuring load balancing, making multi-layer NTN architecture highly scalable and flexible. To further boost performance, several advanced technologies such as RIS, MIMO, NGMA, etc., are being incorporated into NTNs to increase performance.

\textbf{HAPS-SMBS with UAVs mounted RIS}: Integrating HAPS-SMBS with UAVs is an up-and-coming solution to alleviate the constraints of path loss and long distances, however, improving signal propagation remains a critical factor in ensuring network efficiency and reliability. Therefore, the integration of RIS in a multi-layer NTN architecture enhances the network’s ability to dynamically control and redirect signals across its layers. Unlike traditional cooperative techniques such as amplify-and-forward (AF), which have limitations in serving hard-to-reach, high-demand, or congested areas, RIS offers a more efficient solution by strategically directing electromagnetic waves to improve signal propagation and quality \cite{alfattani2021link}. In single-layer systems, RIS is typically deployed on fixed structures, such as building facades, limiting its adaptability. However, in multi-layer systems, RIS mounted on UAVs provide better signal redirection and reduced path loss in complex environments.
This flexibility enables the network to adapt well to fluctuating user demands and changing environmental conditions, thus allowing substantial improvement in coverage, capacity, and overall performance in very complex and harsh environments.
For instance, in areas where the network coverage is poor due to difficult terrain, UAV-RIS dynamically helps in the redirection of signals from HAPS-SMBS to ground devices, hence improving communication quality with reduced loss.
Depending on the structure of the multi-layer NTN architecture, using a two-dimensional (2D) reflective RIS onboard a UAV or other NTN platforms may not be a good option because it only covers one side (i.e., 180 degrees).
To address this issue, a transmissive RIS can be installed on UAVs referred to as UAV-TRIS, as shown in Fig. 1 (b), so that the desired coverage may be achieved by letting the signals pass through the RIS structure. Implementing this structure is an excellent option for NTN communications, providing better services than placing the RIS on one side only.
The installation of the TRIS system on the UAV faces some challenges, such as the airflow from the propeller and aerodynamic issues. Depending on the structure of the UAV and the objective of the TRIS, the latter can be installed in several ways, e.g., on top or on the sides of the UAV as shown in Fig. 2. RIS can be fabricated as an ultra-thin, lightweight, and flexible configuration that could be easily mounted on UAVs with minimized aerodynamic drag and energy consumption. For example, the TRIS can be installed on the top of UAVs to avoid aerodynamic constraints, so that we can utilize the TRIS properly as shown in Fig. 2. 

\begin{figure}[]
    \centering
      \includegraphics[width=\columnwidth]{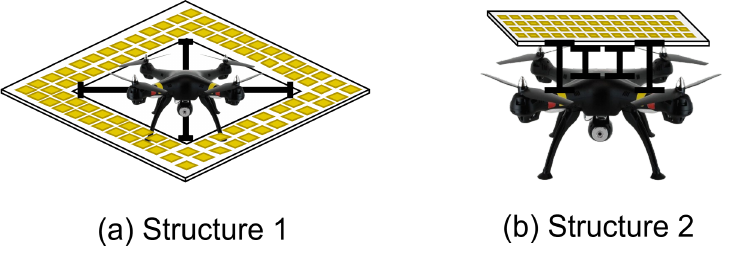}
    \caption{Architecture of TRIS-equipped UAV.}
    \label{constellations}
\end{figure}

\textbf{Multi-layer NTN architecture with Multiple Antennas and NGMA}:
MIMO technologies in HAPS-SMBS systems increase terrestrial device service link capacity, while TAS technology (as illustrated in Fig. 1) reduces the complexity of operating multiple antennas simultaneously, improving service quality and power efficiency. Moreover, incorporating massive MIMO, UM-MIMO, and beamforming deployment enhances capacity and coverage~\cite{abbasi2024hemispherical,10387578}. For instance, UM-MIMO technology can provide reliable communication links between HAPS-SMBS, UAV-TRIS, and ground devices in certain difficult scenarios, such as urban and remote areas. It can be used to form hundreds of high-capacity beams where multiple ground users are simultaneously served. This highlights that MIMO-based spatial diversity contributes to the enhancement of multi-layer NTN architecture that will guarantee superior capacity and reliability relative to single-layer systems.

On the other hand, most studies use OMA to avoid multiple access interference, but it may not be efficient due to NTN's wide coverage and the large number of devices expected. NGMA (e.g., NOMA, RSMA, SDMA, etc) is a promising spectrum-efficient technology for future networks.
Integrating an NGMA system into a multi-layer NTN architecture can provide numerous advantages, including increased energy efficiency, connectivity, and ultra-low latency. 
For example, RSMA adopts advanced power control and signal processing techniques to manage channel conditions between HAPS-SMBS and UAV-TRIS and between UAV-TRIS and ground devices for seamless connectivity across layers ensuring seamless connectivity in very challenging environments. Integration of MIMO and NGMA in multi-layer NTN systems empowers the network to handle high user densities with variable traffic demands; hence, it offers high scalability and adaptability for diverse use cases.

\textbf{Hardware impairments in multi-layer NTN architecture}:
Cost-effective RF transceiver design for mass-market applications usually involves the use of non-ideal components, which are normally prone to manufacturing defects, aging, and environmental conditions. These limitations give rise to RF impairments such as phase noise (PN), direct current (DC) offset, in-phase and quadrature-phase imbalance (IQI), and power amplifier (PA) non-linearity causing in-band and out-of-band interference. These impairments are further aggravated in NTN systems owing to the harsh environmental conditions that greatly degrade the performance of the systems. For example, the sensitivity of PN and IQI might be higher in NTN scenarios compared to conventional terrestrial communications due to large distances, hard weather, and the use of high-frequency bands. The general impairment model is often used to capture the overall effects, thus avoiding the need to evaluate the individual ones.

Moreover, the multi-layer NTN architecture presents a transformative approach to integrating aerial and stratospheric platforms, such as HAPS-SMBS and UAVs, with cutting-edge technologies like TRIS, MIMO, and NGMA. This architecture shows considerable advantages compared to a single layer due to higher coverage, better resource allocation in time, and connectivity resilience. That will truly enable multi-layer systems to exploit completely different complementary strengths, with different layers adapting to multiple scenarios characterized by diverse user demands and environmental conditions. 
As illustrated in Fig. 1 (b) and Table~\ref{tab:compare}, this multi-layer NTN architecture (e.g. HAPS-SMBS and UAVs) with integration of technologies like TRIS, MIMO, and NGMA provides high system capacity, enhanced security, low path-loss, low latency, broader coverage, better energy efficiency, which is in line with the IMT-2030 targets \cite{alfattani2021link,kurt2021vision,9365009,beddiaf2022unified}. 
Despite these advantages and improvements, the system suffers from RF impairments, which may significantly reduce the overall performance. Most studies ignore these impairments in transceivers and assume ideal conditions, which may not provide a complete overview of the overall performance of the systems.

Realistic conditions, such as weather challenges and long transmission distances, further worsen the RF impairments, especially in inter-layer communication. For example, PN effects are more serious while operating at high-frequency bands, such as millimeter wave (mmWave) or terahertz (THz) over a long distance, which poses a serious challenge for accurate beamforming in MIMO systems. While technologies like TRIS, MIMO, and NGMA can alleviate some of these impairments by enhancing the quality of the received signal, their interaction with multi-layer NTN architectures is not well explored.

In order to fill this gap, research into the effects of RF impairments on multi-layer NTN systems is needed. To the best of the authors' knowledge, this work represents the first attempt to thoroughly investigate such effects. The clarity brought by the understanding of how hardware impairments and NTN architectures interact will yield practical methods for mitigation that enhance their practical viability.

\section*{Case Study}
Considering the above-mentioned approach to multi-layer NTN architecture for terrestrial networks, this section presents a case study exploring its potential with RATs under hardware impairments.
The considered system consists of one HAPS-SMBS with $M=4$ antennas, one UAV-TRIS with $N=50$ elements, and $L=3$ terrestrial devices, which are served using NOMA.
It is also assumed that HAPS-SMBS acts as a transmitter (i.e., as a base station), and UAV-TRIS acts as a relay. 
Terrestrial devices receive both direct signals from HAPS-SMBS and indirect signals relayed through UAV-TRIS. To capture the combined effect of RF impairments, hardware impairments are assumed to be present in all nodes as in~\cite{beddiaf2022unified}.
It is also assumed that channel state information is available at all nodes. The model is evaluated under the assumption of a Rician distribution channel.
Since platforms operate at varying altitudes and may thus encounter different attenuation phenomena, we employ realistic path-loss models as specified by 3GPP \cite{3gpp2018study,alfattani2021link}, where path-loss of the links from HAPS-SMBS to UAV and HAPS-SMBS to terrestrial devices are considered to have LoS and Non-LoS (NLoS) conditions. Additionally, the path-loss of the link UAV to terrestrial devices is typically taken to have full LoS conditions. Thus, we consider an urban environment, and the parameters used in all simulations can be found in~\cite{3gpp2018study,alfattani2021link}.

\begin{figure}[]
    \centering
      \includegraphics[width=8cm,height=5.5cm]{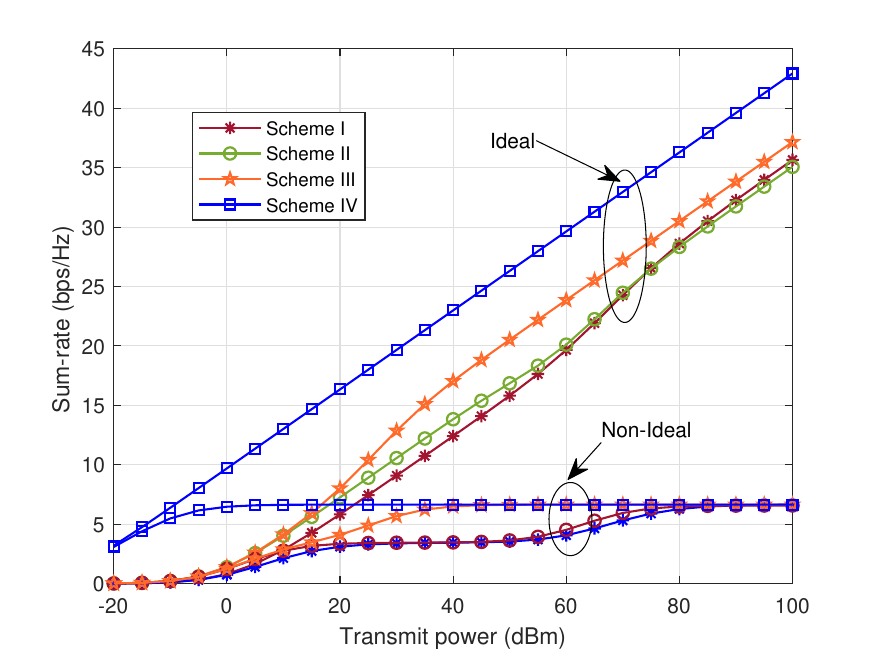}
    \caption{Sum rate comparison of the four schemes with ideal and non-ideal conditions.
    %, when $N=50$ and $M=4$.
    }
    \label{constellations}
\end{figure}

As benchmarks, the proposed system is compared with four schemes as follows. 

\textit{Scheme I} consists of one HAPS-SMBS equipped with a single antenna with direct communication to terrestrial devices.

\textit{Scheme II} includes one HAPS-SMBS with a TAS and one UAV that utilizes an AF protocol. 

\textit{Scheme III} includes one HAPS-SMBS with a TAS and one UAV-mounted passive TRIS.

\textit{Scheme IV} includes one HAPS-SMBS with a TAS and one UAV-mounted active TRIS.

In Fig. 3, we present the sum rate comparison of the four schemes under ideal and non-ideal conditions (i.e., with and without the effect of hardware impairments) with respect to transmit power. 
Scheme II outperforms Scheme I, while Schemes III and IV achieve higher performance gains, but performance deteriorates in non-ideal cases. Despite the use of multiple technologies, hardware impairments significantly limit the achievable performance in multi-layer NTN.

\begin{figure}[]
    \centering
      \includegraphics[width=8cm,height=5.5cm]{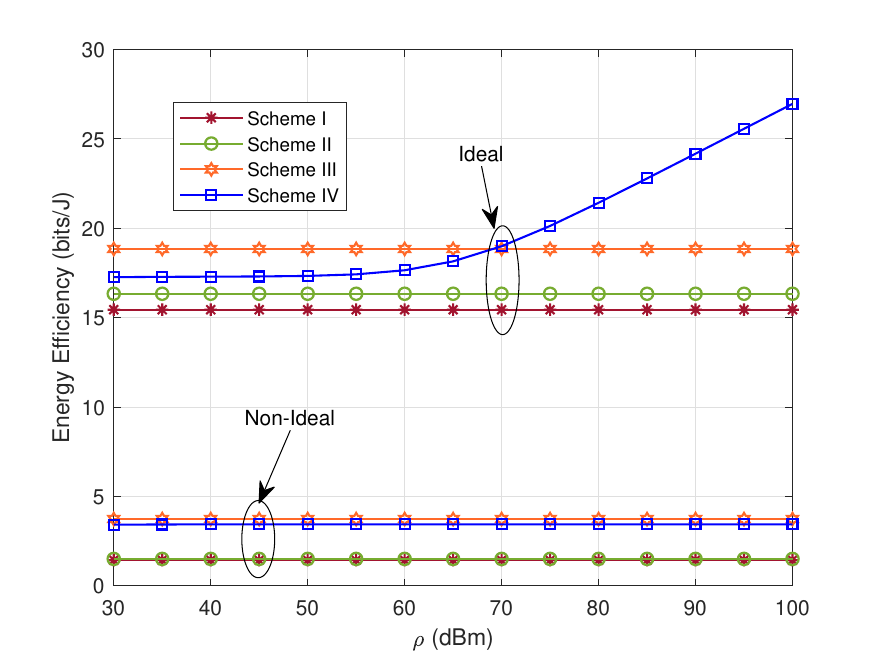}
\caption{Energy efficiency comparison among four schemes with respect to RIS amplitude coefficient ($\rho$) under ideal and non-ideal conditions.
%, when $N=50$ and $M=4$.
}
    \label{constellations}
\end{figure}

On the other hand, the energy efficiency comparison among four schemes with respect to the RIS amplitude coefficient ($\rho$) under ideal and non-ideal conditions when $N=50$ is illustrated in Fig. 4. 
It is observed that increasing the RIS amplitude coefficient significantly enhances the energy efficiency performance of Scheme IV. 
Similarly, Scheme III outperforms Schemes I and II.
However, it can be easily seen that the performance drops dramatically under non-ideal conditions. 

To demonstrate the coverage capabilities of the proposed system, Fig. 5 shows the coverage probability of four schemes with ideal and non-ideal conditions.
We observe that Schemes III and IV have better coverage probabilities than Schemes I and II, while Scheme IV offers superior coverage. However, non-ideal conditions deteriorate the coverage probability performance, with hardware impairments affecting Schemes I and II more due to natural conditions and distance.

From these discussions, the following can be concluded: Multi-layer NTN achieves better performance gains compared to the HAPS-SMBS model. The incorporation of technologies like RIS and MIMO improves coverage and energy efficiency. Active TRIS combats double path-loss and signal amplification but consumes more energy. Hardware impairments significantly deteriorate system performance. Technologies like MIMO and TRIS can reduce hardware impairments through improved signal quality, but mitigation techniques are required.
Moreover, the general impairments (i.e., hardware impairments) model offers a comprehensive performance evaluation framework with NTN systems but requires a future detailed analysis of individual impairments, such as PN, IQI, etc.

\section*{Prospective Use Cases in Multi-Layer NTN Framework}
The multi-layer NTN architecture has the potential to revolutionize connectivity and computing in a variety of contexts. We briefly describe a few potential applications below.

\textbf{ Support for Remote and Inaccessible Areas}:
Traditional terrestrial networks are expensive and difficult to implement and deploy, which calls for integrating NTNs, especially to serve remote areas~\cite{kurt2021vision}. Multi-layer NTNs with advanced technologies can increase coverage, reach inaccessible areas, and enhance network performance. These networks offer vital connectivity for search and rescue operations and real-time data exchange in disaster-stricken areas.
For instance, it would be difficult to construct ground infrastructure in places characterized by mountains or desert regions. In this regard, HAPS can be used to cover a large area, while deployed UAVs fill in the coverage gaps and/or temporary connectivity in emergencies, making the service continuous over areas where single-layer systems (e.g., HAPS) cannot guarantee service.

\begin{figure}[]
    \centering
      \includegraphics[width=8cm,height=5.5cm]{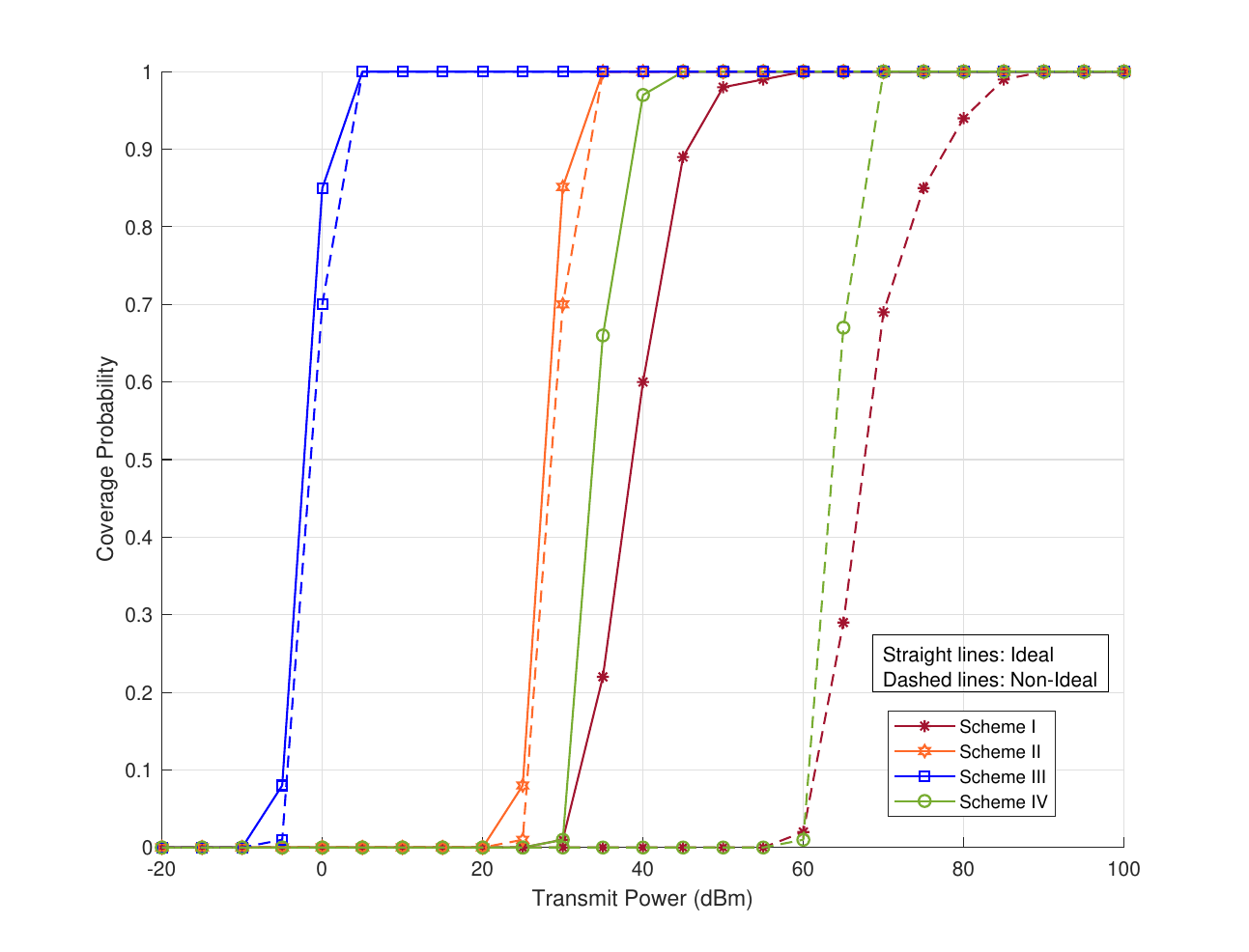}
    \caption{Coverage probability comparison among four schemes with respect to transmit power with ideal and non-ideal conditions.
    %when $N=50$ and $M=4$.
    }
    \label{constellations}
\end{figure}

\textbf{ Support for the Internet of Things}:
The growing adoption of Internet of Things (IoT) technologies, which allow devices to connect to the Internet and exchange data, create difficulties satisfying latency, connectivity, and reliability requirements~\cite{kurt2021vision,9579401}. The integration of multi-layer NTN with IoT could revolutionize connectivity, resource allocation, and emergency operations in urban and industrial environments.
For example, the HAPS systems can enable the collection of massive amounts of data by IoT devices over vast areas, whereas UAVs facilitate mission-critical applications of IoT requiring localized connectivity, such as smart agriculture and industrial automation.
This approach enhances terrestrial networks and supports single-layer systems by enabling seamless and cost-effective connections for IoT data collection.

\textbf{ Enabling Edge Computing}:
Multi-layer NTN with advanced technologies provides a promising solution for mobile edge computing networks by overcoming constraints, such as fixed infrastructure, unavailability in disaster-stricken areas, and low signal strength, enabling flexible computing load distribution, expanded coverage, and improved support for advanced applications~\cite{kurt2021vision}. 
Moreover, in a multi-layer NTN, HAPS might play the role of a central point of edge computing, while UAVs distribute edge computational resources over hot spot areas or across areas with poor infrastructure~\cite{kurt2021vision}.

Despite these capabilities, hardware impairments significantly reduce performance.

\section*{Open Questions, Challenges and Future Research Directions}
This section discusses future challenges in multi-layer NTN, including RF impairment mitigation, UAV energy management, RIS, and antenna designs.

\textbf{ RF/FSO Impairments Mitigation and Compensation}:
Unlike RF systems, free-space optical (FSO) communications in multi-layer NTNs offer high data rates and LoS over an unlicensed spectrum. However, hardware impairments affect signal quality, necessitating comprehensive analysis for compensation solutions~\cite{9503397}. Therefore, in the following years, researchers need to develop specific techniques to compensate for RF/FSO impairments, such as machine learning-based adaptive filtering. Detailed studies could include advanced equalization techniques to mitigate individual impairments like PN and IQI. 

\textbf{ UAV Energy Management: Balancing Flight and Communication}:
We propose several solutions for UAV power management, especially when active TRIS is used. These include sustainable energy sources, charging batteries at landing spots, hybrid active and passive TRIS, trajectory and power control optimization, and UAV-TRIS swarms~\cite{9998527,9579401}. Active UAV-TRIS can save energy and extend service life by remaining fixed on buildings, charging, and continuing communication. On the other hand, future studies can explore hybrid power sources and intelligent trajectory optimization to balance energy consumption. New frameworks could integrate artificial intelligence (AI)-driven energy harvesting and redistribution systems for prolonged UAV operation.

\textbf{ Antenna and RIS Design}:
In NTN systems, meteorological conditions and distance impact service quality, necessitating advanced antenna design for improved satellite and HAPS systems. 
In NTN, ultra-massive MIMO (UM-MIMO) and holographic antenna systems could revolutionize capacity and spectral efficiency, though designing these for NTN platforms’ unique challenges (e.g., stratospheric conditions) remains an open area of research. Hybrid antenna systems that combine traditional MIMO with adaptive beamforming can be explored to optimize coverage and reduce interference. Additionally, holographic MIMO antennas could pave the way for spatial multiplexing on a massive scale, enabling extensive connectivity. However, RF impairment issues persist in NTN networks~\cite{abbasi2024hemispherical,10232975}. RIS design is important in NTN communications; therefore, 3D RIS or 2D TRIS technology may provide better coverage. Efforts could also focus on miniaturized, lightweight antenna systems optimized for stratospheric platforms.

\textbf{ Security in Multi-layer NTN architecture}:
Advanced technologies in multi-layer NTN can enhance security by enhancing confidentiality rates, weakening terrestrial eavesdropping attempts, and amplifying legitimate signals. However, secure communication between layers and ground users is crucial for data-oriented applications. UAV jammers and AI can offer promising solutions for continuous secure communication~\cite{kurt2021vision,9579401}. Future systems can employ quantum cryptography in NTN communications to protect against emerging threats. This could include post-quantum cryptographic algorithms resilient to quantum attacks.

\textbf{ Interference Management in Multi-layer NTN}:
NGMA techniques like NOMA, RSMA, and SDMA improve spectral efficiency and support massive connectivity. However, interference caused by overlapping signals in multi-layer NTN environments is a critical challenge. Advanced interference management algorithms, powered by AI, can dynamically optimize spectrum use~\cite{10397567}. The integration of NGMA with machine learning can enhance resource allocation and reduce latency. Techniques like deep reinforcement learning could predict user demand patterns, enabling proactive interference management and efficient spectrum sharing~\cite{kurt2021vision,10397567}.

\textbf{ Handover Management in Multi-layer NTN}:
Handover management of multi-layer NTN is complex in dense urban areas due to signal quality and latency fluctuations, which affect network stability and user experiences. 
Therefore, we should focus on designing robust algorithms for seamless user transitions between NTN and/or terrestrial layers. AI-based predictive models could proactively manage handover decisions based on user mobility and network conditions~\cite{10397567}. This should also include multi-layer coordination strategies, i.e., developing frameworks for efficient coordination among layers (e.g., satellite, HAPS, and UAV)~\cite{kurt2021vision}. These frameworks could optimize resource sharing, reduce redundancy, and improve overall system efficiency.

\textbf{Standardization Efforts:} Standardization plays a vital role in ensuring that multi-layer NTN systems operate seamlessly across different platforms and regions. Leading organizations like ITU and 3GPP have made significant progress in defining guidelines for integrating NTNs into 6G frameworks. For example, ITU IMT-2030 emphasizes the need for NTN systems to complement terrestrial networks, while 3GPP Release 18 introduces specific adaptations for satellite communication and high-altitude platforms. Future efforts should prioritize the development of a unified protocol for NTN and terrestrial systems to enhance interoperability and scalability. Moreover, aligning these efforts with global initiatives such as the UN Sustainable Development Goals (e.g., SDG 9: Industry, Innovation, and Infrastructure) can accelerate widespread adoption and ensure the technology's societal impact.

\textbf{Environmental Impact Assessments:} The deployment of NTNs comes with environmental challenges, including the carbon footprint of satellite launches and the energy consumption of high-altitude platforms. Incorporating renewable energy sources like solar panels for HAPS and UAVs can mitigate some of these issues. Furthermore, using recyclable materials in NTN components and adopting modular designs for UAVs can reduce electronic waste. These measures align with SDG 13 (Climate Action) by promoting sustainable practices. Additionally, the ecological effects of NTN operations on atmospheric and terrestrial ecosystems require further study, with solutions like AI-driven environmental monitoring helping to minimize disruptions. Sustainable deployment strategies will ensure that NTNs support global connectivity without compromising environmental health.

\section*{Conclusion}
This article has focused on exploring the potential of a multi-layer NTN architecture subject to hardware impairments. 
The architecture includes a HAPS-SMBS, UAVs equipped with either passive or active TRIS, NGMA, MIMO, and terrestrial devices. 
Afterward, potential application scenarios under hardware impairments are presented and discussed.
Additionally, we suggest some challenges and potential solutions to overcome hardware impairments and other operational limitations in NTN.
Simulation results show that hardware impairments significantly limit the capabilities of NTN systems.
Therefore, effective modeling, optimization, and dynamic compensation are required for the successful implementation and improvement of future communication systems.

\section*{Acknowledgment}
This study is supported in part by the Ministry of Higher Education and Scientific Research of Algeria (MESRS), in part by The Scientific and Technological Research Council of Türkiye (TUBITAK), and in part by the Discovery Grant RGPIN-2022-05231 from the Natural Sciences and Engineering Research Council of Canada (NSERC).

\bibliographystyle{ieeetr}

\end{document}